\documentclass[aps,preprint]{revtex4}
\usepackage{amsmath}
\usepackage{mathrsfs}
\usepackage{bm}

\begin{document}

\title{The three-wave coupling coefficients for a cold magnetized plasma}
\author{L. Stenflo and G. Brodin \\
%EndAName
Department of Physics, Ume\aa\ University, SE-901 87 Ume\aa , Sweden}

\begin{abstract}
The resonant interaction between three waves in a uniform magnetized plasma
is considered. Using the somewhat inaccessible result of the general theory
we deduce the explicit expressions for the coupling coefficients of a cold
magnetized two-component plasma.
\end{abstract}

\maketitle

\draft

%\pacs{52.35.Mw}

\section{Introduction}

The physics of nonlinear waves is a rapidly developing research field which
has recently received increased attention (e.g. Shukla 2004; Stenflo 2004;
Onishchenko \textit{et al 2004}; Vladimirov and Yu 2004; Wu and Chao 2004;
Shukla and Stenflo 2005a,b; Azeem and Mirza 2005; Mendonca \textit{et al.
2005; }Marklund and Shukla 2005). Although there are general formalisms to
treat such phenomena, there is also a need to have access to reliable
explicit expressions for specific cases. As a particular example, Brodin and
Stenflo (1988; 1990) considered the resonant interaction between three MHD
waves in a plasma. Starting from the standard MHD theory they thus derived
the coupling coefficients. However, the textbook MHD equations are not able
to correctly treat the nonlinear interaction between three Alfv\'{e}n waves
(c.f. Shukla and Stenflo 2005b). In the present paper we are therefore going
to reconsider the general nonlinear interaction between three waves in a
cold, magnetized, two-component plasma, in order to derive the \textit{%
explicit }expressions for the coupling coefficients. Such expressions have
previously been presented for a one-component plasma (Stenflo 1973), but,
due to algebraic difficulties, never before for a two-component plasma.

\section{Results}

Considering the resonant interaction between three waves with frequencies $%
\omega _{j}$ ($j=1,2,3$) and wave vectors $\mathbf{k}_{j}$, we assume that
the matching conditions 
\begin{equation}
\omega _{3}=\omega _{1}+\omega _{2}  \label{frequency}
\end{equation}
and 
\begin{equation}
\mathbf{k}_{3}=\mathbf{k}_{1}+\mathbf{k}_{2}  \label{wave-vector}
\end{equation}
are satisfied. The development of, for example, the z-components ($E_{jz})$
of the wave electric field amplitudes is then governed by the three coupled
bilinear equations (e.g. Stenflo 1994) 
\begin{mathletters}
\begin{equation}
\frac{dE_{1z}^{\ast }}{dt}=\alpha _{1}E_{2z}E_{3z}^{\ast }  \tag{3a}
\end{equation}
\begin{equation}
\frac{dE_{2z}^{\ast }}{dt}=\alpha _{2}E_{1z}E_{3z}^{\ast }%
\addtocounter{equation}{2}  \tag{3b}
\end{equation}
and 
\begin{equation}
\frac{dE_{3z}}{dt}=\alpha _{3}E_{2z}E_{1z}  \tag{3c}
\end{equation}
where the z-axis is along the external magnetic field ($B_{0}\widehat{%
\mathbf{z}}$), the star denotes complex conjugate, $\alpha _{j}$ are the
coupling coefficients, $d/dt=\partial /\partial t+\mathbf{v}_{gj}\cdot
\nabla +\nu _{j}$ where $\mathbf{v}_{gj}$ is the group velocity of wave $j$,
and $\nu _{j}$ accounts for the linear damping rate. The general formula for 
$\alpha _{j}$ for a \textit{hot} magnetized plasma has been derived
previously (e.g. Stenflo and Larsson 1977; Stenflo 1994) and it is therefore
here only presented in Appendix A. Although, in principle, it covers all
interaction mechanisms in uniform plasmas, it is not easy to apply it
directly to, for example, Alfv\'{e}n waves in a \textit{cold} two-component
plasma. However, after some straightforward, but rather lengthy evaluation
of the formula in Appendix A, we can finally write $\alpha _{j}$ in the
comparatively simple form 
\end{mathletters}
\begin{equation}
\alpha _{1,2}=\frac{M_{1,2}}{\partial D(\omega _{1,2},\mathbf{k}%
_{1,2})/\partial \omega _{1,2}}C  \tag{4a,b}
\end{equation}
and 
\begin{equation}
\alpha _{3}=-\frac{M_{3}}{\partial D(\omega _{3},\mathbf{k}_{3})/\partial
\omega _{3}}C  \tag{4c}
\end{equation}
where 
\begin{eqnarray}
C &=&\sum_{\sigma }\frac{q\omega _{p}^{2}}{m\omega _{1}\omega _{2}\omega
_{3}k_{1z}k_{2z}k_{3z}}\left[ \frac{\mathbf{k}_{1}\mathbf{\cdot K}_{1}}{%
\omega _{1}}\mathbf{K}_{2}\mathbf{\cdot K}_{3}^{\ast }+\frac{\mathbf{k}_{2}%
\mathbf{\cdot K}_{2}}{\omega _{2}}\mathbf{K}_{1}\mathbf{\cdot K}_{3}^{\ast }+%
\frac{\mathbf{k}_{3}\mathbf{\cdot K}_{3}^{\ast }}{\omega _{3}}\mathbf{K}_{1}%
\mathbf{\cdot K}_{2}-\right.   \notag \\
&&\left. \frac{i\omega _{c}}{\omega _{3}}\left( \frac{k_{2z}}{\omega _{2}}-%
\frac{k_{1z}}{\omega _{1}}\right) \mathbf{K}_{3}^{\ast }\mathbf{\cdot }%
\left( \mathbf{K}_{1}\times \mathbf{K}_{2}\right) \right]   \label{C-coeff}
\end{eqnarray}
\begin{equation}
\mathbf{K}=-\left[ \mathbf{k}_{\bot }+i\frac{\omega _{c}}{\omega }\mathbf{k}%
\times \widehat{\mathbf{z}}+\left( \frac{\sum i\frac{\omega _{c}}{\omega }%
\frac{\omega _{p}^{2}}{\omega ^{2}-\omega _{c}^{2}}}{1-\frac{k^{2}c^{2}}{%
\omega ^{2}}-\sum \frac{\omega _{p}^{2}}{\omega ^{2}-\omega _{c}^{2}}}%
\right) \left( \mathbf{k}\times \widehat{\mathbf{z}}-i\frac{\omega _{c}}{%
\omega }\mathbf{k}_{\bot }\right) \right] \frac{\left( 1-\frac{k_{\bot
}^{2}c^{2}}{\omega ^{2}}-\sum \frac{\omega _{p}^{2}}{\omega ^{2}}\right)
\omega ^{4}}{\left( \omega ^{2}-\omega _{c}^{2}\right) k_{\bot }^{2}c^{2}}%
+k_{z}\widehat{\mathbf{z}}  \label{k-coeff-vector}
\end{equation}

\begin{eqnarray}
D(\omega ,\mathbf{k}) &=&\left( 1-\frac{k^{2}c^{2}}{\omega ^{2}}-\sum \frac{%
\omega _{p}^{2}}{\omega ^{2}-\omega _{c}^{2}}\right) \left[ \left( (1-\frac{%
k_{z}^{2}c^{2}}{\omega ^{2}}-\sum \frac{\omega _{p}^{2}}{\omega ^{2}-\omega
_{c}^{2}}\right) \left( 1-\frac{k_{\bot }^{2}c^{2}}{\omega ^{2}}-\sum \frac{%
\omega _{p}^{2}}{\omega ^{2}}\right) -\right.   \notag \\
&&\left. \frac{k_{\bot }^{2}k_{z}^{2}c^{4}}{\omega ^{4}}\right] -\left( \sum 
\frac{\omega _{p}^{2}\omega _{c}}{\omega (\omega ^{2}-\omega _{c}^{2})}%
\right) ^{2}\left( 1-\frac{k_{\bot }^{2}c^{2}}{\omega ^{2}}-\sum \frac{%
\omega _{p}^{2}}{\omega ^{2}}\right)   \label{DR}
\end{eqnarray}
and 
\begin{equation}
M_{j}=\left( 1-\frac{k_{j}^{2}c^{2}}{\omega _{j}^{2}}-\sum \frac{\omega
_{p}^{2}}{\omega _{j}^{2}-\omega _{c}^{2}}\right) \left( 1-\frac{%
k_{jz}^{2}c^{2}}{\omega _{j}^{2}}-\sum \frac{\omega _{p}^{2}}{\omega
_{j}^{2}-\omega _{c}^{2}}\right) -\left( \sum \frac{\omega _{p}^{2}\omega
_{c}}{\omega _{j}(\omega _{j}^{2}-\omega _{c}^{2})}\right) ^{2}
\label{M-formula}
\end{equation}
where $k=(k_{z}^{2}+k_{\bot }^{2})^{1/2}$, $\mathbf{k}_{\bot }$ is the
perpendicular (to $\widehat{\mathbf{z}}$) part of the wave-vector, $\omega
_{p}$ is the plasma frequency ($\omega _{pe}$ for the electrons and $\omega
_{pi}$ for the ions), $\omega _{c}=qB_{0}/m$ is the cyclotron frequency, $q$
and $m$ are the particle charge and mass, and $c$ is the speed of light in
vacuum. For notational convenience, the subscript $\sigma $ denoting the
various particle species has been left out in the above formulas. We stress
that no approximations have to be used to derive the expressions (4)-(8)
which thus are quite general for the case of three wave interactions in a
cold magnetized two-component plasma. It can also be verified that (4)
agrees with the coupling coefficients for a magnetized one-component
(Stenflo 1973; 1994) plasma.

Equations (3a,b,c), with (4), significantly improve the (approximate)
equations in the previous work by Brodin and Stenflo (1988) for the case
when the plasma is cold. Thus, although the main emphasis in that work was
on the coupling between Alfv\'{e}n waves and magnetosonic waves where useful
results were derived, it was also mentioned that there is no coupling
between Alfv\'{e}n waves in the MHD limit. The present paper shows however
that this is not true. Thus there is a non-zero interaction between, for
example, one dispersive Alfv\'{e}n pump wave (Shukla and Stenflo 2005b) and
two inertial Alfv\'{e}n waves characterized by 
\begin{equation}
\omega _{1,2}\simeq \frac{k_{1,2z}V_{A}}{1+k_{1,2\bot }^{2}\lambda _{e}^{2}}
\label{inertial}
\end{equation}
where $V_{A}$ is the Alfv\'{e}n velocity and $\lambda _{e}=c/\omega _{pe}$.
In the particular case when $E_{z}$ is zero for one of the waves, it is of
course straightforward to use other variables, e.g. $E_{x}$ instead of $E_{z}
$, to derive expressions similar to those above.

\section{Conclusions}

In the present paper we have improved the approximate results for three wave
interactions in an MHD plasma (Brodin and Stenflo 1988) and found the 
\textit{explicit }expressions for the coupling coefficients for wave
interactions in a cold magnetized two-component plasma. Our coupling
coefficient $C$ can thus be used as a starting point (see for example
Appendix B) of any estimate of the coupling strength where the interaction
between any kind of waves (Alfv\'{e}n waves, whistler waves, etc.) in a cold
plasma has to be considered. It can also be useful in interpretations of
stimulated scattering of electromagnetic waves in space plasmas (e.g.
Stenflo 1999; Kuo 2001; 2003). In the latter case we refer the reader to a
short historical account of stimulated electromagnetic emissions in the
ionosphere (Stenflo 2004).

\section{Appendix A}

When calculating the coupling coefficients, it turns out that they contain a
common factor $\mathrm{V}$. It is then possible to write the three coupled
equations as 
\begin{equation}
\frac{dW_{1,2}}{dt}=-2\omega _{1,2}\mathrm{{Im}V}  \tag{A1}
\end{equation}
and 
\begin{equation}
\frac{dW_{3}}{dt}=2\omega _{3}\mathrm{{Im}V}  \tag{A2}
\end{equation}
where $W=\varepsilon _{0}\mathbf{E}^{\ast }\cdot (1/\omega )\partial (\omega
^{2}\bm{\varepsilon })\mathbf{E}$ is the wave energy $\bm{
\varepsilon }$ is the usual textbook dielectric tensor, and $\mathrm{{Im}V}$
stands for the imaginary part of $\mathrm{V}$ where (Stenflo and Larsson
1977; Stenflo 1994) 
\begin{equation*}
\mathrm{V}=\sum_{s}m\int d\mathbf{vF}_{0}(\mathbf{v})\sum_{\substack{ %
p_{1}+p_{2}=p_{3} \\ p_{j}=0,\pm 1,\pm 2,...}}%
I_{1}^{p_{1}}I_{2}^{p_{2}}I_{3}^{-p_{3}}\left[ \frac{\mathbf{k}_{1}\mathbf{%
\cdot u}_{1p_{1}}}{\omega _{1d}}\mathbf{u}_{2p_{2}}\mathbf{\cdot u}%
_{3p_{3}}^{\ast }+\frac{\mathbf{k}_{2}\mathbf{\cdot u}_{2p_{2}}}{\omega _{2d}%
}\mathbf{u}_{1p_{1}}\mathbf{\cdot u}_{3p_{3}}^{\ast }+\frac{\mathbf{k}_{3}%
\mathbf{\cdot u}_{3p_{3}}^{\ast }}{\omega _{3d}}\mathbf{u}_{1p_{1}}\mathbf{%
\cdot u}_{2p_{2}}\right. 
\end{equation*}
\begin{equation}
\left. -\frac{i\omega _{c}}{\omega _{3d}}\left( \frac{k_{2z}}{\omega _{2d}}-%
\frac{k_{1z}}{\omega _{1d}}\right) \mathbf{u}_{3p_{3}}^{\ast }\mathbf{\cdot }%
\left( \mathbf{u}_{1p_{1}}\times \mathbf{u}_{2p_{2}}\right) \right]  
\tag{A3}
\end{equation}
where $\omega _{jd}=\omega _{j}-k_{jz}v_{z}-p_{j}\omega _{c}$, $I_{j}$ ($%
=\exp (i\theta _{j})$)$=(k_{j_{x}}+ik_{jy})/k_{j\bot }$, and the velocity $%
\mathbf{u}_{jp_{j}}$ satisfies 
\begin{equation*}
\omega _{jd}\mathbf{u}_{jp_{j}}+i\omega _{c}\widehat{\mathbf{z}}\times 
\mathbf{u}_{jp_{j}}=\frac{iq}{m\omega _{j}}\left\{ \omega _{jd}J_{p_{j}}%
\mathbf{E}_{j}+\left[ \left( v_{z}E_{jz}+\frac{p_{j}\omega _{c}}{k_{j\bot
}^{2}}\mathbf{k}_{j\bot }\mathbf{\cdot E}_{j\bot }\right) J_{p_{j}}+\right.
\right. 
\end{equation*}
\begin{equation}
\left. \left. \frac{iv_{\bot }\omega _{c}}{k_{j\bot }^{2}}(\widehat{\mathbf{z%
}}\times \mathbf{k}_{j})\mathbf{\cdot E}_{j}\frac{d}{dv_{\bot }}J_{p_{j}}%
\right] \mathbf{k}_{j}\right\}   \tag{A4}
\end{equation}
where $J_{p_{j}}=J_{p_{j}}(k_{j\bot }v_{\bot }/\omega _{c})$ denotes a
Bessel function of order $p_{j}$.

\section{Appendix B}

The limit when $\omega $ is much smaller than $\omega _{ci}$ is of special
interest. In that case, we approximate (6) by 
\begin{equation}
\mathbf{K}_{e}\mathbf{\simeq }-\frac{i\omega }{\omega _{ce}}\frac{(1+k_{\bot
}^{2}\lambda _{e}^{2})}{k_{\bot }^{2}\lambda _{e}^{2}}\mathbf{k}\times 
\widehat{\mathbf{z}}+k_{z}\widehat{\mathbf{z}}  \tag{B1}
\end{equation}
and 
\begin{equation}
\mathbf{K}_{i}\mathbf{\simeq }-\frac{i\omega }{\omega _{ci}}\frac{(1+k_{\bot
}^{2}\lambda _{e}^{2})}{k_{\bot }^{2}\lambda _{e}^{2}}\left[ \mathbf{k}%
\times \widehat{\mathbf{z}}-\frac{i\omega }{\omega _{ci}}\mathbf{k}_{\bot }%
\right]  \tag{B2}
\end{equation}
We note that the ion contributions dominate the first three terms in (5),
whereas the electron contributions are most important for the fourth term in
(5). As a result we have a non-zero coupling coefficient $C$ for the
particular case of interaction between three Alfv\'{e}n waves, in contrast
to what one obtains from the too simplified textbook MHD equations (Brodin
and Stenflo 1988; where $C_{AAA\text{ }}$was zero).

\end{document}